\documentclass[12pt]{article}
\begin{document}
\begin{center}
{\large\bf Dynamical Constraints in the Nonsymmetric Gravitational Theory}
\vskip 0.3 true in
{\large J. W. Moffat}
\vskip 0.3 true in
{\it Department of Physics, University of Toronto,
\vskip 0.3 true in
Toronto, Ontario, Canada M5S 1A7}
\end{center}
\begin{abstract}%
We impose in the nonsymmetric gravitational theory, by means
of Lagrange multiplier fields in the action, a set of covariant constraints on the
antisymmetric tensor field. The canonical Hamiltonian constraints in the
weak field approximation for the antisymmetric sector yield a Hamiltonian
energy bounded from below. An analysis of the Cauchy evolution, in terms of
an expansion of the antisymmetric sector about a symmetric Einstein
background, shows that arbitrarily small antisymmetric Cauchy data can lead
to smooth evolution.
\end{abstract}

\section{Introduction}

Recently, a modified version of the nonsymmetric gravitational theory (NGT)
field equations has been
published~\cite{Moffat1,Moffat2,Moffat3,Moffat4,Clayton1}, which leads to a
linear approximation for weak fields that incorporates Einstein's linear
field equations and the massive Kalb-Ramond field equations for the
antisymmetric field $g_{[\mu\nu]}$~\cite{van,Ramond,Moffat5}. These
equations do not have any ghost poles, tachyons or unphysical asymptotic
behaviour, removing the inconsistencies discovered by Damour, Deser and
McCarthy, in an earlier version of NGT~\cite{Damour}.
Clayton~\cite{Clayton2,Clayton3,Clayton4} has developed a Hamiltonian
formalism for NGT that reveals a local instability of the NGT field
equations associated with three of the six possible propagating degrees of
freedom in the skew sector. Thus, the three components $g_{[0i]}$
$(i,j=1,2,3)$ of the six components $g_{[\mu\nu]}$ become unphysical
propagating degrees of freedom. Since NGT does not possess a rigorous gauge
invariance associated with the skew sector of the form:
\begin{equation}
\label{transform} g_{[\mu\nu]}\rightarrow
g_{[\mu\nu]}+\xi_{\mu,\nu}-\xi_{\nu,\mu},
\end{equation}
where $\xi_\mu$ is
an arbitrary vector field, we must impose dynamical constraints on the
action through covariant Lagrangian constraint equations, which guarantee
that the antisymmetric transverse-longitudinal degrees of freedom vanish
throughout the evolution of the dynamical NGT field equations. We shall
show in the following how this can be implemented and how the new field
degrees of freedom associated with the Lagrange multiplier fields produce
consistent field equations in the linear approximation and a Hamiltonian
bounded from below.

We also show that the Cauchy instability deduced by
Clayton~\cite{Clayton2,Clayton3,Clayton4} for the massive
NGT field equations is no longer present in the evolution of the equations for
an expansion of the antisymmetric sector about an arbitrary Einstein background.

\section{The NGT Action and the Constrained Field Equations}

We shall decompose the nonsymmetric $g_{\mu\nu}$ and
$\Gamma^\lambda_{\mu\nu}$ as
\begin{equation}
g_{(\mu\nu)}={1\over 2}(g_{\mu\nu}+g_{\nu\mu}),\quad g_{[\mu\nu]}=
{1\over 2}(g_{\mu\nu}-g_{\nu\mu}),
\end{equation}
and
\begin{equation}
\Gamma^\lambda_{\mu\nu}=\Gamma^\lambda_{(\mu\nu)}
+\Gamma^\lambda_{[\mu\nu]}.
\end{equation}
The contravariant tensor $g^{\mu\nu}$ is defined in terms of the equation:
\begin{equation}
\label{inverse}
g^{\mu\nu}g_{\sigma\nu}=g^{\nu\mu}g_{\nu\sigma}={\delta^\mu}_\sigma.
\end{equation}
The notation follows that of earlier work on
NGT~\cite{Moffat1,Moffat2,Moffat3,Moffat4}.
The Lagrangian density is given by
\begin{equation}
{\cal L}_{ngt}={\cal L}+{\cal L}_M,
\end{equation}
where
\begin{eqnarray}
\label{NGTLagrangian}
{\cal L}={\bf g}^{\mu\nu}R_{\mu\nu}(W)-2\Lambda\sqrt{-g}
-{1\over 4}\mu^2{\bf g}^{\mu\nu}g_{[\nu\mu]}\nonumber\\
-{1\over 6}{\bf g}^{\mu\nu}
W_\mu W_\nu
+{\bf g}^{\mu\nu}J_{[\mu}\phi_{\nu]},
\end{eqnarray}
and ${\cal L}_M$ is the matter Lagrangian density ($G=c=1$):
\begin{equation}
{\cal L}_M=-8\pi g^{\mu\nu}{\bf T}_{\mu\nu}.
\end{equation}
Here, ${\bf g}^{\mu\nu}=\sqrt{-g}g^{\mu\nu}$, $g=\hbox{Det}(g_{\mu\nu})$,
$\Lambda$ is the cosmological constant and
$R_{\mu\nu}(W)$ is the NGT contracted curvature tensor:
\begin{equation}
R_{\mu\nu}(W)=W^\beta_{\mu\nu,\beta}
- {1\over 2}(W^\beta_{\mu\beta,\nu}+W^\beta_{\nu\beta,\mu}) -
W^\beta_{\alpha\nu}W^\alpha_{\mu\beta} +
W^\beta_{\alpha\beta}W^\alpha_{\mu\nu},
\end{equation}
defined in terms of the unconstrained nonsymmetric connection:
\begin{equation}
\label{Wequation}
W^\lambda_{\mu\nu}=\Gamma^\lambda_{\mu\nu}-{2\over 3}{\delta^\lambda}_\mu
W_\nu,
\end{equation}
where
\[
W_\mu={1\over 2}(W^\lambda_{\mu\lambda}-W^\lambda_{\lambda\mu}).
\]
Eq.(\ref{Wequation}) leads to the result:
\[
\Gamma_\mu=\Gamma^\lambda_{[\mu\lambda]}=0.
\]
The contracted tensor $R_{\mu\nu}(W)$ can be written as
\[
R_{\mu\nu}(W)=R_{\mu\nu}(\Gamma)+\frac{2}{3}W_{[\mu,\nu]},
\]
where
\[
R_{\mu\nu}(\Gamma ) = \Gamma^\beta_{\mu\nu,\beta} -{1\over 2}
\left(\Gamma^\beta_{(\mu\beta),\nu} + \Gamma^\beta_{(\nu\beta),\mu}\right) -
\Gamma^\beta_{\alpha\nu} \Gamma^\alpha_{\mu\beta} +
\Gamma^\beta_{(\alpha\beta)}\Gamma^\alpha_{\mu\nu}.
\]

The term in Eq.(\ref{NGTLagrangian}):
\begin{equation}
\label{Multiplier}
{\bf g}^{\mu\nu}J_{[\mu}\phi_{\nu]},
\end{equation}
contains the Lagrange multiplier fields $\phi_\mu$ and the source vector
$J_\mu$.

A variation of the action
\[
S=\int d^4x{\cal L}_{\hbox{ngt}}
\]
yields the field equations in the presence of matter sources:
\begin{equation}
\label{Gequation}
G_{\mu\nu} (W)+\Lambda g_{\mu\nu}+S_{\mu\nu}
=8\pi (T_{\mu\nu}+K_{\mu\nu}),
\end{equation}
\begin{equation}
\label{divg}
{{\bf g}^{[\mu\nu]}}_{,\nu}=-\frac{1}{2}{\bf g}^{(\mu\alpha)}W_\alpha,
\end{equation}
\begin{equation}
{{\bf g}^{\mu\nu}}_{,\sigma}+{\bf g}^{\rho\nu}W^\mu_{\rho\sigma}
+{\bf g}^{\mu\rho}
W^\nu_{\sigma\rho}-{\bf g}^{\mu\nu}W^\rho_{\sigma\rho}
+{2\over 3}\delta^\nu_\sigma{\bf g}^{\mu\rho}W^\beta_{[\rho\beta]}
$$ $$
+{1\over 6}({\bf g}^{(\mu\beta)}W_\beta\delta^\nu_\sigma
-{\bf g}^{(\nu\beta)}W_\beta\delta^\mu_\sigma)=0.
\end{equation}
Here, we have $G_{\mu\nu}=R_{\mu\nu} - {1\over 2}g_{\mu\nu}R$ and
\begin{equation}
S_{\mu\nu}=\frac{1}{4}\mu^2(g_{[\mu\nu]}
+{1\over 2}g_{\mu\nu}g^{[\sigma\rho]}
g_{[\rho\sigma]}+g^{[\sigma\rho]}g_{\mu\sigma}g_{\rho\nu})
$$ $$
-\frac{1}{6}(W_\mu W_\nu-\frac{1}{2}g_{\mu\nu}g^{\alpha\beta}W_\alpha W_\beta).
\end{equation}

Moreover, the contribution from the variation of (\ref{Multiplier})
with respect to $g^{\mu\nu}$ and $\sqrt{-g}$ is given by
\begin{equation}
\label{Kequation}
K_{\mu\nu}=-\frac{1}{8\pi}[J_{[\mu}\phi_{\nu]}-\frac{1}{2}
g_{\mu\nu}(g^{[\alpha\beta]}J_{[\alpha}\phi_{\beta]})].
\end{equation}

The variation of $\phi_\mu$ yields the constraint equations
\begin{equation}
\label{gskewconstraint}
{\bf g}^{[\mu\nu]}J_\nu=0.
\end{equation}
These equations hold globally for the evolution of the field equations.  We have
not varied the source vector $J_\mu$.

If we use (\ref{gskewconstraint}), then (\ref{Kequation}) becomes
\begin{equation}
\label{K2}
K_{[\mu\nu]}=-\frac{1}{8\pi}J_{[\mu}\phi_{\nu]}.
\end{equation}

We can choose the vector $J_\mu$ to be $J_\mu=(0,0,0,J_0)$, so
that (\ref{gskewconstraint}) corresponds to the three constraint equations
\begin{equation}
\label{constraints2} {\bf g}^{[i0]}=0.
\end{equation}

The generalized Bianchi identities
\begin{equation}
[{\bf g}^{\alpha\nu}G_{\rho\nu}(\Gamma)+{\bf g}^{\nu\alpha}
G_{\nu\rho}(\Gamma)]_{,\alpha}+{g^{\mu\nu}}_{,\rho}{\bf G}_{\mu\nu}=0,
\end{equation}
give rise to the matter response equations~\cite{Moffat4}:
\begin{equation}
g_{\mu\rho}{{\bf T}^{\mu\nu}}_{,\nu}+g_{\rho\mu}{{\bf T}^{\nu\mu}}_{,\nu}
+(g_{\mu\rho,\nu}+g_{\rho\nu,\mu}-g_{\mu\nu,\rho}){\bf T}^{\mu\nu}=0.
\end{equation}

\section{Linear Approximation}

Let us assume that $\Lambda=0$ and expand $g_{\mu\nu}$ about Minkowski
spacetime:
\begin{equation}
g_{\mu\nu}=\eta_{\mu\nu}+{}^{(1)}h_{\mu\nu}+...,
\end{equation}
where $\eta_{\mu\nu}$ is the Minkowski metric tensor: $\eta_{\mu\nu}=
\hbox{diag}(-1, -1, -1, +1)$. We shall also expand $\Gamma^\lambda_{\mu\nu}$,
$W^\lambda_{\mu\nu}$ and $K_{[\mu\nu]}$:
\begin{eqnarray}
\Gamma^\lambda_{\mu\nu}&=&{}^{(1)}\Gamma^\lambda_{\mu\nu}
+{}^{(2)}\Gamma^\lambda_{\mu\nu}+...,\\
W^\lambda_{\mu\nu}&=&{}^{(1)}W^\lambda_{\mu\nu}
+{}^{(2)}W^\lambda_{\mu\nu}+...,\\
K_{[\mu\nu]}&=&{}^{(1)}K_{[\mu\nu]}+{}^{(2)}K_{[\mu\nu]}+... .
\end{eqnarray}

We adopt the notation, $\psi_{\mu\nu}={}^{(1)}h_{[\mu\nu]}$, and using
(\ref{inverse}), we find that
$\psi^{\mu\nu}=\eta^{\mu\lambda}\eta^{\sigma\nu}\psi_{\sigma\lambda}$. To first
order of approximation, Eq.(\ref{divg}) gives
\begin{equation}
\label{psiequation}
\psi_\mu=-{1\over 2}W_\mu,
\end{equation}
where for convenience $W_\mu$ denotes ${}^{(1)}W_\mu$. Moreover,
\[
\psi_\mu={\psi_{\mu\beta}}^{,\beta}=\eta^{\beta\sigma}\psi_{\mu\beta,\sigma}.
\]

The antisymmetric and symmetric field equations derived from Eq.(\ref{Gequation})
decouple to lowest order and the symmetric equations are the usual
Einstein field equations in the linear approximation. The skew
equations are given by~\cite{Moffat2}
\begin{equation}
\label{Procaequation}
{\psi_{\mu\nu,\sigma}}^{,\sigma}+\mu^2\psi_{\mu\nu}=J_{\mu\nu},
\end{equation}
where
\begin{equation}
J_{\mu\nu}=16\pi(T_{[\mu\nu]}
+{2\over \mu^2}{T_{[[\mu\sigma],\nu]}}^{,\sigma}+
K_{[\mu\nu]}+{2\over\mu^2}{K_{[[\mu\sigma],\nu]}}^{,\sigma}).
\end{equation}

We can write the Lagrangian density for the $\psi_{\mu\nu}$ in the form:
\begin{equation}
\label{skewLagrangian}
{\cal L}_{\hbox{skew}}=\frac{1}{12}F_{\mu\nu\lambda}F^{\mu\nu\lambda}
-\frac{1}{4}\mu^2\psi_{\mu\nu}\psi^{\mu\nu}
-16\pi\psi^{\mu\nu}(T_{[\mu\nu]}+K_{[\mu\nu]}),
\end{equation}
where
\[
F_{\mu\nu\lambda}=\psi_{\mu\nu,\lambda}+\psi_{\nu\lambda,\mu}
+\psi_{\lambda\mu,\nu},
\]
and to linear order we have from Eq.(\ref{K2}):
\begin{equation}
\label{Kaequation}
K_{[\mu\nu]}=-\frac{1}{8\pi}J_{[\mu}\phi_{\nu]},
\end{equation}
where for convenience, we have used $K_{[\mu\nu]}$
in place of ${}^{(1)}K_{[\mu\nu]}$ and $\phi_\mu$ in place of
${}^{(1)}\phi_\mu$. From Eq.(\ref{psiequation}) we get
\begin{equation}
\label{divK}
\psi_\mu=\frac{16\pi}{\mu^2}({T_{[\mu\nu]}}^{,\nu}+{K_{[\mu\nu]}}^{,\nu}).
\end{equation}
The source current density $T_{[\mu\nu]}$ is identified at
the microscopic level with a bosonic open string current
density~\cite{Ramond,Moffat5}, and for open strings the $T_{[\mu\nu]}$ is
not conserved.

In the wave-zone, $T_{\mu\nu}=0$, and Eqs. (\ref{Procaequation}) and
(\ref{divK}) become
\begin{eqnarray}
\label{Boxequation}
{\psi_{\mu\nu,\sigma}}^{,\sigma}-{\psi_{\mu\sigma,\nu}}^{,\sigma}
+{\psi_{\nu\sigma,\mu}}^{,\sigma} +\mu^2\psi_{\mu\nu}&=&16\pi
K_{[\mu\nu]},\\
\label{gauge}
\psi_\mu&=&\frac{16\pi}{\mu^2}{K_{[\mu\nu]}}^{,\nu}.
\end{eqnarray}
Eq.(\ref{Boxequation}) can be written as
\begin{equation}
{F_{\mu\nu\sigma}}^{,\sigma}+\mu^2\psi_{\mu\nu}=16\pi K_{[\mu\nu]}.
\end{equation}

By using the frame in which $J_\mu=(0,0,0,J_0)$ and substituting the weak
field constraint, $\psi_{i0}=0$, into (\ref{gauge}), we obtain
\begin{equation}
\label{Bdiv}
\phi_{i,i}+J_{0,i}\phi_i=0,
\end{equation}
and
\begin{equation}
\label{aequation}
\label{dotaequation}
{\dot \phi}_i=-\mu^2(\psi_{ij,j}+\frac{1}{\mu^2}{\dot J}_0\phi),
\end{equation} where ${\dot \phi}_i =\partial \phi_i/\partial t$.

Choosing $\mu=0$ and $\nu=i$ in (\ref{Boxequation}), we get
\begin{equation}
\label{dotpsiequation}
{\dot\psi}_{ij,j}=J_0\phi_i,
\end{equation}
while choosing $\mu=i$ and $\nu=k$ yields an equation for the transverse-transverse
components of $\psi_{\mu\nu}$:
\begin{equation}
\label{Boxpsi}
{\psi_{ik,\sigma}}^{,\sigma}-{\psi_{ij,k}}^{,j}
+{\psi_{kj,i}}^{,j}+\mu^2\psi_{ik}=0,
\end{equation} where we have used
\begin{equation}
K_{[0i]}=-\frac{1}{16\pi}J_0\phi_i,\quad K_{[ik]}=0
\end{equation}
obtained from (\ref{Kaequation}).

By using (\ref{dotaequation}) and the time derivative of (\ref{dotpsiequation}), we get
\begin{equation}
{\ddot\psi}_{ij,j}=-\mu^2\psi_{ij,j}.
\end{equation}
This equation has the solution:
\begin{equation}
\label{cequation}
\psi_{ij,j}({\vec x},t)=c_i({\vec x})\hbox{cos}(\mu t+\theta),
\end{equation}
where $\theta$ is a constant. By taking the divergence of (\ref{Boxpsi}),
we obtain
\begin{equation}
\label{ddotpsi}
{\ddot\psi}_{ij,j}-2\nabla^2\psi_{ij,j}+\mu^2\psi_{ij,j}=0.
\end{equation}
Substituting the solution (\ref{cequation}) into (\ref{ddotpsi}) gives
\begin{equation}
\label{nablac}
\nabla^2c_i({\vec x})=0.
\end{equation}

Let us substitute
\begin{equation}
\psi_{ik}({\vec x},t)=b_{ik}({\vec x})\hbox{cos}(\mu t+\theta)
\end{equation}
into (\ref{Boxpsi}) and use (\ref{cequation}) to give
\begin{equation}
\label{nablab}
\nabla^2b_{ik}({\vec x})=2c_{[i,k]}({\vec x}).
\end{equation}
By employing the solution of the Laplace equation (\ref{nablac}), we can obtain
the $b_{ik}$ from (\ref{nablab}) which in turn determines
the transverse-transverse components $\psi_{ik}$.

For the static case, we obtain from (\ref{dotaequation}):
\[
\psi_{ij,j}=0
\]
and (\ref{Boxpsi}) becomes
\begin{equation}
\nabla^2\psi_{ik}({\vec x})-\mu^2\psi_{ik}({\vec x})=0,
\end{equation}
which is the Helmholtz equation with the point source, spherically symmetric
Yukawa solution:
\begin{equation}
\psi_{ik}(r)=\lambda_{ik}\frac{\hbox{exp}(-\mu r)}{r},
\end{equation}
where $\lambda_{ik}$ and $\mu$ are constants.

\section{Canonical Form of the Hamiltonian in the Weak Field Antisymmetric
Sector}

By decomposing the Lagrangian (\ref{skewLagrangian})
for $T_{[\mu\nu]}=0$ into space and time components and using
$u_\mu=(0,0,0,1)$, we get
\begin{eqnarray}
{\cal L}_{\hbox{skew}}
=\frac{1}{4}[({\dot\psi}_{ik})^2+4\psi_{k0,i}{\dot\psi}_{ik}
+2(\psi_{k0,i})^2-2\psi_{k0,k}\psi_{i0,i}+2\psi_{kj,k}\psi_{ij,i}\nonumber\\
-(\psi_{jk,i})^2-\mu^2((\psi_{ik})^2-2(\psi_{0i})^2)] + J_0\phi_i\psi_{0i}.
\end{eqnarray}
The conjugate momenta $\pi_{\mu\nu}$ are given by
\begin{eqnarray}
\pi_{i0}&=&\frac{\partial{\cal L}_{\hbox{skew}}}{\partial{{\dot\psi}_{i0}}}=0,\\
\pi_{ij}&=&\frac{\partial{\cal L}_{\hbox{skew}}}{\partial{{\dot\psi}_{ij}}}
={\dot\psi}_{ij}+\psi_{0i,j}-\psi_{0j,i}.
\end{eqnarray}
We can now obtain the following form of the Lagrangian:
\begin{eqnarray}
\label{Lequation}
{\cal L}_{\hbox{skew}}=\frac{1}{4}[(\pi_{ij})^2+2\psi_{jk,k}\psi_{ji,i}
-(\psi_{jk,i})^2-\mu^2((\psi_{ik})^2-2(\psi_{0i})^2)]\nonumber\\
+ J_0\phi_i\psi_{0i}.
\end{eqnarray}

The Hamiltonian is given by
\begin{equation}
{\cal H}_{\hbox{skew}}=\frac{1}{2}\pi_{ik}{\dot\psi}_{ik}-{\cal
L}_{\hbox{skew}}
+\theta_i\pi_{i0},
\end{equation}
where $\theta_i$ acts as a Lagrange multiplier for the constraint $\pi_{i0}=0$.
Using (\ref{Lequation}) we get
\begin{eqnarray}
{\cal H}_{\hbox{skew}}=\frac{1}{4}(\pi_{ik})^2+\pi_{ij}\psi_{i0,j}-\frac{1}{2}
\psi_{jk,k}\psi_{ji,i}+\frac{1}{4}(\psi_{jk,i})^2+\frac{1}{4}\mu^2((\psi_{ik})^2
-2(\psi_{i0})^2)\nonumber\\
+J_0\phi_i\psi_{i0}+\theta_i\pi_{i0}.
\end{eqnarray}
By employing the Poisson bracket relation:
\begin{equation}
\{\psi_{i0},\pi_{j0}\}=\delta_{ij},
\end{equation}
we get the evolution equation:
\begin{equation}
\label{pidot}
{\dot\pi}_{i0}=\{\pi_{i0}, H_{\hbox{skew}}\}\approx -\pi_{ik,k}-\mu^2\psi_{i0}
+ J_0\phi_i\approx 0,
\end{equation}
where
\[
H_{\hbox{skew}}=\int d^3x{\cal H}_{\hbox{skew}}.
\]

Varying with respect to $\phi_i$ yields the constraint
\[
\psi_{i0}\approx 0.
\]
Taking the divergence of (\ref{pidot}) leads to Eq.(\ref{Bdiv}).
Moreover, we have
\begin{equation}
{\dot\psi}_{i0}=\{\psi_{i0},H_{\hbox{skew}}\}\approx \theta_i\approx 0.
\end{equation}
We have the set of constraints:
\begin{equation}
\label{6equations}
\pi_{i0}\approx 0,\quad \psi_{i0}\approx 0,
\end{equation}
which constitute six second class constraints.
Thus, there are (6 degrees of freedom)-$\frac{1}{2}$(6 second class constraints)=3
independent degrees of freedom.

Imposing the constraints, we obtain the Hamiltonian energy of the system:
\begin{equation}
H_{\hbox{skew}}\approx \int d^3x\frac{1}{2}[\pi_i^2+(\psi_{i,i})^2
+\mu^2\psi_i^2],
\end{equation}
where $\pi_i=\frac{1}{2}\epsilon_{ijk}\pi_{jk}$ and
$\psi_i=\frac{1}{2}\epsilon_{ijk}
\psi_{jk}$. Thus, the energy is bounded from below and the weak field
antisymmetric sector leads to a physically consistent Hamiltonian, free of ghost
poles and tachyons.

\section{Linearization about a Fixed Einstein Background}

In recent papers, Clayton has developed a Hamiltonian constraint formalism
for NGT~\cite{Clayton2,Clayton3,Clayton4}. The field equations are
decomposed into a (3+1) form by foliating spacetime into spacelike
hypersurfaces. Configuration space has therefore been chosen to consist of
$g^{\perp i}$ and $g^{ik}$, where the index $\perp$ denotes a component
normal to a hypersurface $\Sigma$. The Cauchy data is chosen so that the
antisymmetric sector is an arbitrarily small perturbation of the symmetric
sector. Thus, the Cauchy data is expanded in powers of the antisymmetric
components $g^{[\mu\nu]}=(B^i,\gamma^{[ik]})$ about an arbitrary Einstein
background described by the symmetric metric $\gamma^{(ik)}$.

From the Hamiltonian constraint formalism, equations for the velocity components
${\dot {\bf B}}^i$ and the acceleration components ${\ddot {\bf B}}^i$ are
derived. These are given by (see ref.~\cite{Clayton3,Clayton4} for
details): \begin{eqnarray}
\label{Hamconstraint1} {\dot {\bf B}}^i\approx
Y^i+\frac{3}{4}N\gamma^{(ik)}O^{-1}_{2kj}Z^j,\\ \label{Hamconstraint2}
{\ddot {\bf B}}^i\approx\frac{3}{4}N\sqrt{S}O^{-1ik}_2X_k,
\end{eqnarray}
where \begin{equation} O^{-1}_{2ik}\sim
-\frac{1}{{\vec\gamma}\cdot{\vec\gamma}}\biggl[\gamma_{(ik)}
-\frac{1}{{\vec\gamma}\cdot{\vec B}}(B_i\gamma_k+\gamma_k B_i)\nonumber\\
+\frac{{\vec B}\cdot{\vec{B}}
-{\vec\gamma}\cdot{\vec\gamma}}{({\vec\gamma}\cdot {\vec
B})^2}\gamma_i\gamma_k\biggr].
\end{equation} The $X_i, Y_i$ and $Z_i$ are
quantities that are well behaved as the antisymmetric components become
vanishingly small and ${\bf B}_i$ is the densitised $B_i$. Moreover,
$\gamma_i=\frac{1}{4}\sqrt{S}\epsilon_{ijk}\gamma^{[jk]},
S=\hbox{Det}(S_{ik})$ with $S_{ik}$ defined by
$\gamma^{(ik)}S_{kj}={\delta^i}_j$, $N$ is the lapse function and we have
used the notation: ${\vec A}\cdot{\vec B}=A^i B_i$. The operator
$O^{-1}_{2ik}\sim O(\hbox{skew}^{-2})$ is singular in the limit of
arbitrarily small antisymmetric fields.

In the version of NGT presented here, the global constraints (\ref{gskewconstraint})
and (\ref{constraints2}),
obtained from the action principle, guarantee that the
velocity components ${\dot {\bf B}}^i$ and the acceleration components
${\ddot {\bf B}}^i$,
determined by Eqs.(\ref{Hamconstraint1}) and (\ref{Hamconstraint2}),
are absent from the evolution equations. Thus, this
version of NGT can yield smooth solutions arbitrarily close to the symmetric
Einstein background solutions, allowing for stable Cauchy evolution and
linearization. However, further work is required to demonstrate that
there exists a complete, rigorous solution to the Cauchy evolution problem.

\section{Conclusions}

We have formulated a version of NGT with a covariant action principle, including
constraints on the antisymmetric field variables $g_{[\mu\nu]}$, implemented by
the Lagrange multiplier fields $\phi_\mu$, which guarantees a system of
consistent field equations. By expanding the antisymmetric field sector about
Minkowski spacetime and performing a Hamiltonian constraint analysis, it was
shown that the Hamiltonian for the weak antisymmetric fields was bounded
from below, ensuring a physically consistent weak field approximation in
conjunction with the weak field symmetric Einstein sector.

The field equations were then expressed in the (3+1) Hamiltonian
formalism developed by Clayton~\cite{Clayton2,Clayton3,Clayton4}, and the
antisymmetric sector was expanded about
an arbitrary symmetric Einstein background. It was then shown that the
new field equations with the dynamical constraints, ${\bf
g}^{[\mu\nu]}J_\nu=0$, led to Cauchy stable evolution equations and
linearization stable solutions for arbitrarily small $g_{[\mu\nu]}$ fields.
Therefore, both the massless and the massive NGT field equations yield
physically consistent solutions with asymptotically flat boundary
conditions with well-defined Newtonian and Einstein gravity limits.
\vskip
0.3 true in Acknowledgments
\vskip 0.3 true in I thank P. Savaria, M.
Clayton, L. Demopoulos, J. L\'egar\'e and I. Yu. Sokolov for helpful and
stimulating discussions. This work was supported by the Natural Sciences
and Engineering Research Council of Canada.

\end{document}